\documentclass[preprint, iop]{emulateapj}
\usepackage{graphicx}
\usepackage{amsmath}
\usepackage{natbib}
\usepackage{times}
\usepackage[colorlinks=true,urlcolor=blue,citecolor=blue,linkcolor=blue]{hyperref}
\usepackage{aas_macros}
\newcommand{\ha}{H$\alpha$}

\begin{document}
\title{On the multi-threaded nature of solar spicules}
\author{H. Skogsrud\altaffilmark{1}}
\author{L. Rouppe van der Voort\altaffilmark{1}}
\author{B. De Pontieu\altaffilmark{1,2}}
\affil{\altaffilmark{1}Institute of Theoretical Astrophysics, University of Oslo, P.O. Box 1029 Blindern, N-0315 Oslo, Norway}
\affil{\altaffilmark{2}Lockheed Martin Solar \& Astrophysics Lab, Org.\ A021S, Bldg.\ 252, 3251 Hanover Street Palo Alto, CA~94304 USA}

\begin{abstract}
A dominant constituent in the dynamic chromosphere are spicules.
Spicules at the limb appear as relatively small and dynamic jets that are observed to everywhere stick out.
Many papers emphasize the important role spicules might play in the energy and mass balance of the chromosphere and corona.
However, many aspects of spicules remain a mystery.
In this Letter we shed more light on the multi-threaded nature of spicules and their torsional component.
We use high spatial, spectral and temporal resolution observations from the Swedish 1-m Solar Telescope in the \ha\ spectral line.
The data targets the limb and we extract spectra from spicules far out from the limb to reduce the line-of-sight superposition effect.
We discover that many spicules display very asymmetric spectra with some even showing multiple peaks.
To quantify this asymmetry we use a double Gaussian fitting procedure and find an average velocity difference between the single Gaussian components to be between 20--30~km~s$^{-1}$ for a sample of 57 spicules.
We observe that spicules show significant sub-structure where one spicule consists of many 'threads'.
We interpret the asymmetric spectra as line-of-sight superposition of threads in one spicule and therefore have a measure for a perpendicular flow inside spicules which will be important for future numerical model to reproduce.
In addition we show examples of $\lambda-x$-slices perpendicular across spicules and find spectral tilts in individual threads providing further evidence for the complex dynamical nature of spicules.
\end{abstract}

\section{Introduction}
Spicules are relatively small and highly dynamic jets that protrude everywhere from the limb as observed in chromospheric spectral lines.
Their nature is still a mystery owing to their dynamic nature and small size, ranging from some hundred kilometers all the way down to the resolution limits of state of the art telescopes.
Spicules have been extensively studied in the past and the earliest discovery of spicules dates back to 1877 by Secchi.
The early spicule observations and models are reviewed by \cite{beckers_1968} and \cite{sterling_2000}. 

Using \ion{Ca}{2} H images from the then recently launched Hinode satellite \citep{hinode_overview,sot_overview,sot} \cite{de_pontieu_et_al_2007b} found that spicules come in two classes, type I and type II.
The distinction between type I and type II spicules is based on the striking differences in dynamic behavior in the Ca II H 3968 \AA\ passband of Hinode/SOT.
Type II spicules are the dominant type in quiet-Sun and coronal holes regions while type I spicules are almost exclusively found in active regions. 
Type I spicules show a slower apparent outward velocity (15-40~km~s$^{-1}$) and are seen to fall back toward the limb after reaching maximum height, whereas type II spicules show faster outward velocity (30-110~km~s$^{-1}$) and are observed to fade from the Ca II H passband at their maximum extent likely due to heating. 
While it is unclear what drives type I spicules, their dynamic behavior show similarities with those of dynamic fibrils in active region and some quiet Sun mottles which have been studied in detail by \cite{hansteen_et_al_2006}, \cite{de_pontieu_et_al_2007a}, \cite{rouppe_van_der_voort_et_al_2007} and \cite{martinez-sykora_et_al_2009}.
%Type I spicules are almost exclusively found in active regions and they are generally less dynamic.
%Their lifetimes are longer and they show consecutive upward and downward motion.
%Type I spicules have been extensively studied both observationally and numerically, see \cite{hansteen_et_al_2006} and \cite{de_pontieu_et_al_2007a} with references therein.
%Type II spicules tend to disappear at their maximum extent likely due to heating.
Hints of sub-structure in spicules have been found in Hinode data, \cite{suematsu_et_al_2008} and \cite{sterling_et_al_2010} suggest that many spicules have two components.

\cite{de_pontieu_et_al_2011} show that brightenings appear in hotter passbands in the Atmospheric Imaging Assembly (AIA, \cite{lemen_et_al_2012}) after type II spicules disappear from \ha, which strengthens their importance as energy mediators in the outer atmosphere.
It is however not clear to what extend spicules reach coronal temperatures or how their energy is deposited in the corona. 

There are a very limited amount of numerical studies of type II spicules compared to the wealth of observational data that exists.
\cite{martinez-sykora_et_al_2011} study in detail one spicule similar to a type II spicule that evolved naturally in their large scale simulation.
Type II spicules are generally not present in large scale simulations of the solar atmosphere, even though observations indicate that type II spicules are the dominant constituent in the chromosphere and corona interface region.

The motion of type II spicules is found to consist of a combination of tree motions: Field-aligned flows, transverse swaying of the central axis and torsional rotation around the central axis.
This is observed from spicules off-limb  \citep{de_pontieu_et_al_2007a,pereira_2012,de_pontieu_et_al_2012_twist} and from the disk-counterpart of spicules (rapid blue-shifted excursion) \citep{langangen_et_al_2008b,rouppe_van_der_voort_et_al_2009,sekse_et_al_2012, sekse_et_at_2013_temporal,sekse_et_al_2013_3motions}.

The aim of this paper is to study the multi-threaded nature of spicules in high resolution images and the torsional component in more detail by searching for clean spectra of spicules far off-limb.
Targeting off-limb is especially important because the spectra are not as optically thick compared to on-disk and therefore are easier to interpret. 
Given the enormous line-of-sight effects close to the limb we need to go to great heights to reduce superposition.

\section{Observations}
We analyze a high quality dataset obtained with the CRISP instrument \citep{scharmer_et_al_2008_crisp} at the Swedish 1-m Solar Telescope (SST, \cite{scharmer_et_al_2003}).
CRISP is an imaging spectro-polarimeter using dual Fabry-P\'erot etalons as described by \cite{scharmer_2006} and it works as a tunable narrowband filter that can rapidly scan spectral lines isolated by a prefilter.
One camera is situated just after the prefilter called the wide band camera (WB) and two more cameras are placed after the dual etalons.
In \ha\ the full width at half maximum of the transmission profile for CRISP is 6.6~pm and 0.49~nm for the \ha\ prefilter.
 
The data analyzed was from 27 June 2010 between 11:43:12 $-$ 12:25:25~UTC.
The adaptive optics system \citep{scharmer_et_al_2003a} was locking on a region of enhanced network about 30\arcsec\ inside from the limb.
The pixel size is 0.059\arcsec\ and the entire field of view (FOV) covers about 57$\times$57\arcsec\ of the Sun.
Both observations from STEREO \citep{secchi} from the same day and AIA a few days later show that the solar region below the off-limb part is magnetically less active than the enhanced network in the center of the FOV.
The size of the FOV region off-limb is approximately $100$~Mm$^2$.

The observational sequence scanned the \ha\ line at 41 equidistant positions symmetrically around line-center.
The scan started at $-1716$~m\AA\ with steps of $86$~m\AA, equivalent to a Doppler offset of $-80$~km~s$^{-1}$ in steps of $4$~km~s$^{-1}$.
The cadence of the series is 9~s and 281 scans were recorded.
 
The data targeted a region of plage just inside the south-eastern limb of the Sun.
A sunspot was situated close to, but outside the FOV.

The observations were post-processed with the MOMFBD \citep{van_noort_et_al_2005} procedure.
The MOMFBD procedure assures that the NB images are all aligned to the WB image and therefore to each other, thereby achieving a high degree of spectral integrity in the spectral scan.
The WB and NB objects are given equal weight in the restoration process.
This process may leave residual seeing artifacts due to changes in the differential seeing.
This is compensated for by adding another object to the restoration following \cite{henriques_2012}.

It is important to note that off-limb there is reduced signal in the wide-band channel.
This makes it difficult to compute the wavefront deformation and makes the restoration less successful.
However, the seeing in this dataset is excellent and very stable and it was verified that spurious signals from spatial misalignment due to the sequential nature of the acquisition method are negligible.

We used CRISPEX \citep{vissers_et_al_2012} to explore and analyze the observations.

\section{Analysis}
\begin{figure}[!t]
  \centering
  \includegraphics[width=\columnwidth]{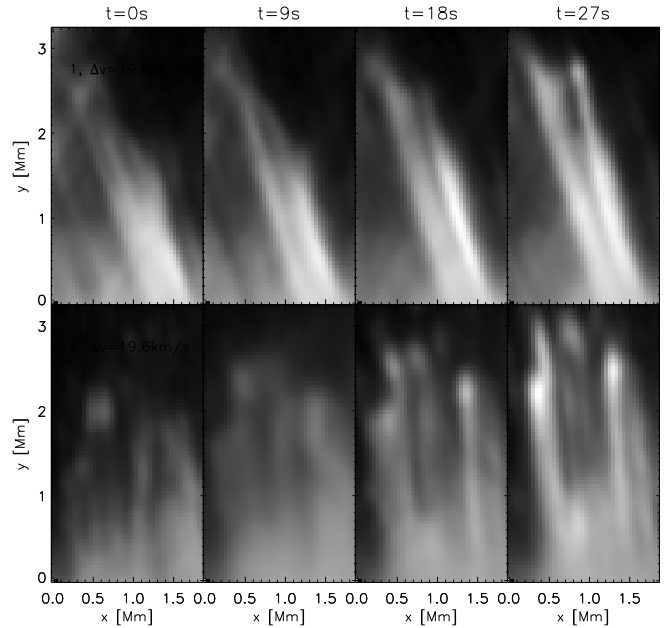}
  \caption{Time evolution of two spatial regions each containing several spicules from \ha\ red-wing images ($\Delta v=19.6$~km~s$^{-1}$).
    The images are radial filtered to make the spicules stand out more and the cutouts are fixed in space.
    [Animation of this figure is available in the online material].}
  \label{fig:tfig}
\end{figure}

\begin{figure}[!t]
  \centering
  \includegraphics[width=0.8\columnwidth]{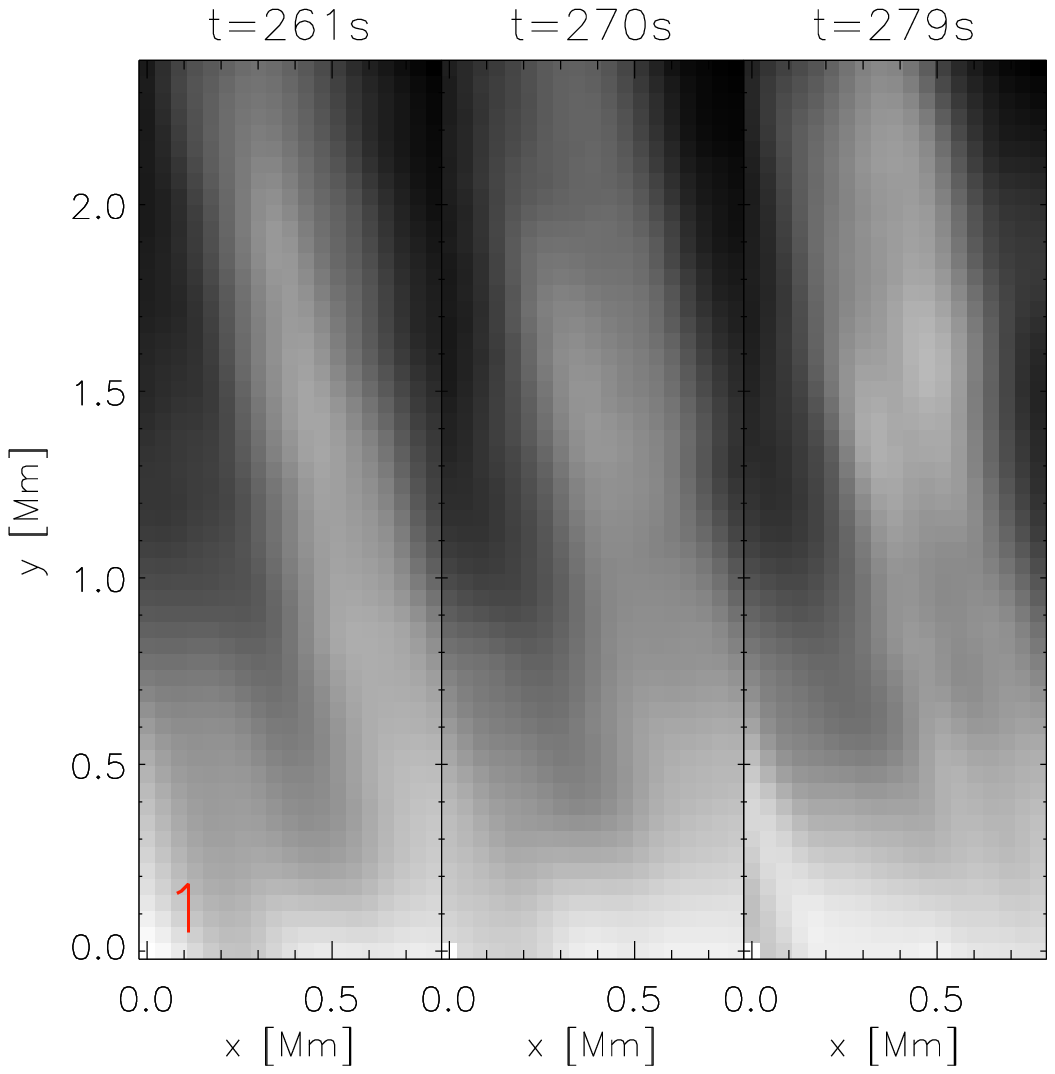}
  \includegraphics[width=0.8\columnwidth]{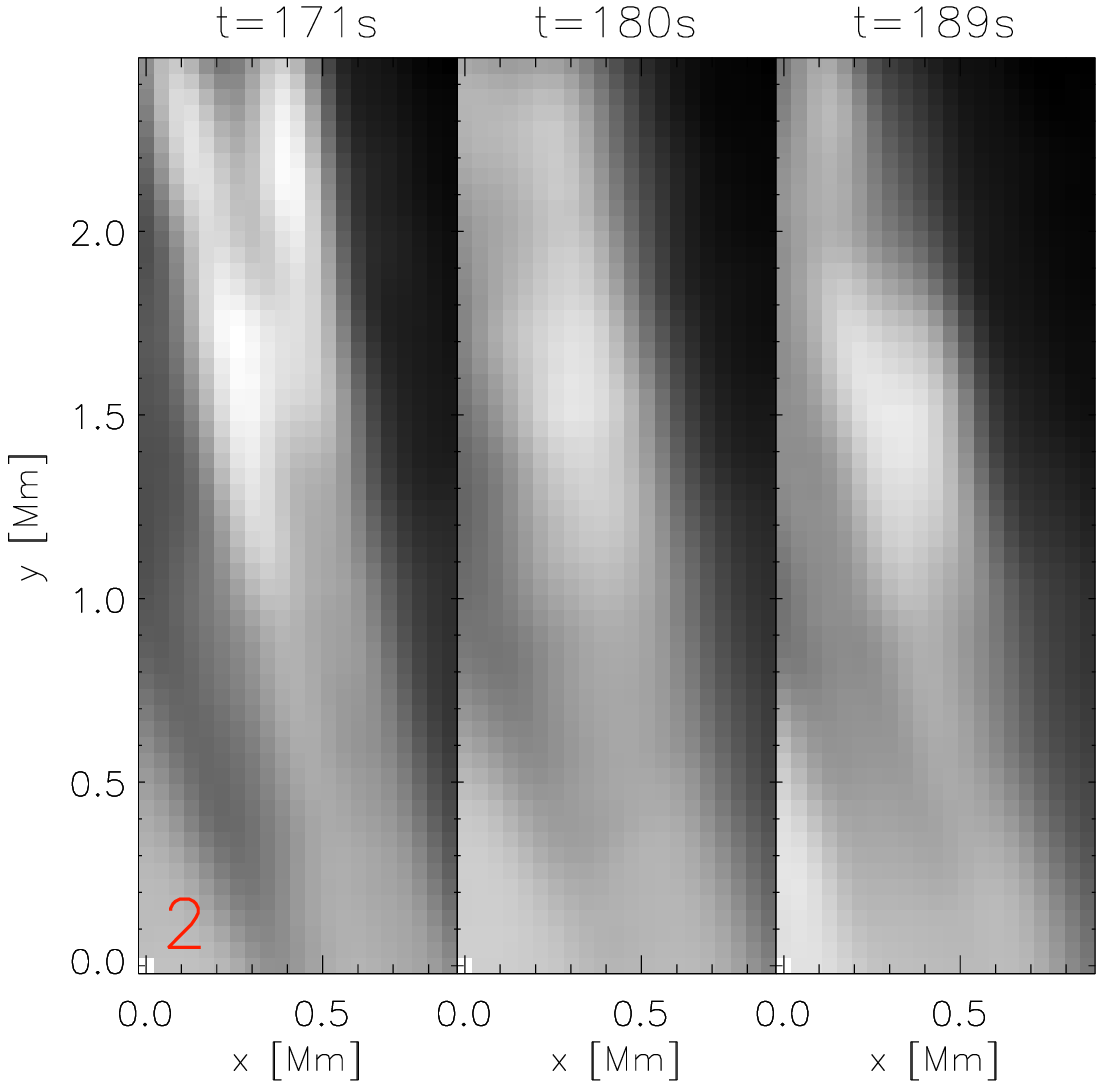}
  \includegraphics[width=0.8\columnwidth]{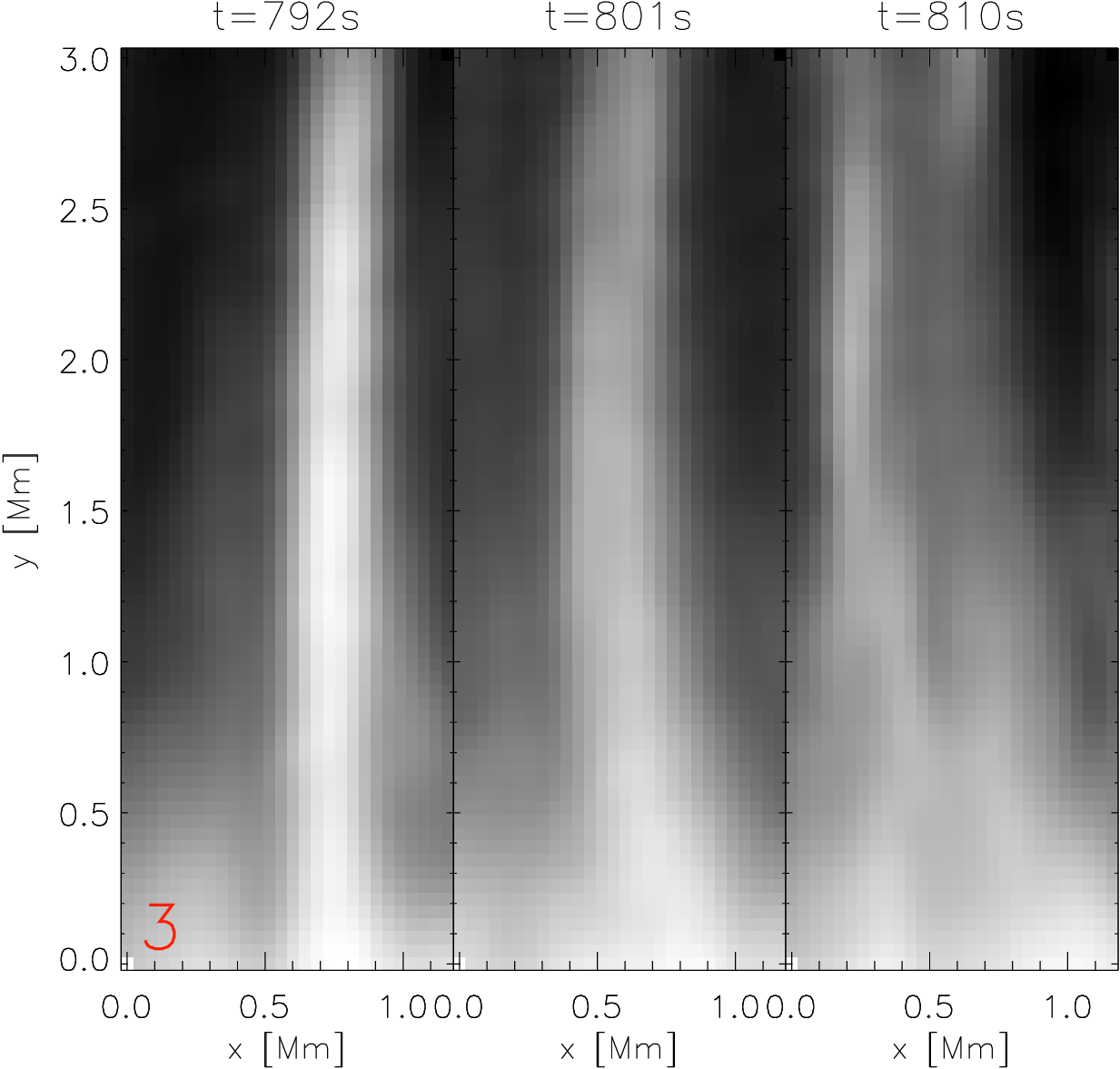}
  \caption{Three examples of spicules splitting and merging in a short period of time.
    The images are from \ha\ line center and are radial filtered to make the spicules stand out more.
    The cutouts are fixed in space.}
  \label{fig:split}
\end{figure}

When we look at the data it is apparent that many spicules do not evolve independently, but rather evolve as small groups of spicules in a collective fashion.
The spicules in the groups appear and disappear nearly at the same time and have similar apparent motion.
Two examples are shown in Fig.~\ref{fig:tfig}.
In the first example two spicules have an apparent outward motion.
After 81~s (not shown in the figure, see movie in the online material) the rightmost spicule splits into two spicules.
After an additional 27~s the leftmost spicule splits into two spicules.
In the second example a larger ensemble of spicules have appeared and appears to move outward.
The very bright spicule to the left at $t=27$~s appear to be made up of two spicules and at later times they appear to separate.
In Fig.~\ref{fig:split} we show three examples of spicules merging or splitting up over a short period of time.
In the first and third example the splitting appears along nearly the entire length of the spicule at the same time indicating that the relative motion of the individual spicules is perpendicular to their axis.
 
\begin{figure*}[!t]
  \centering
  \includegraphics[width=\textwidth]{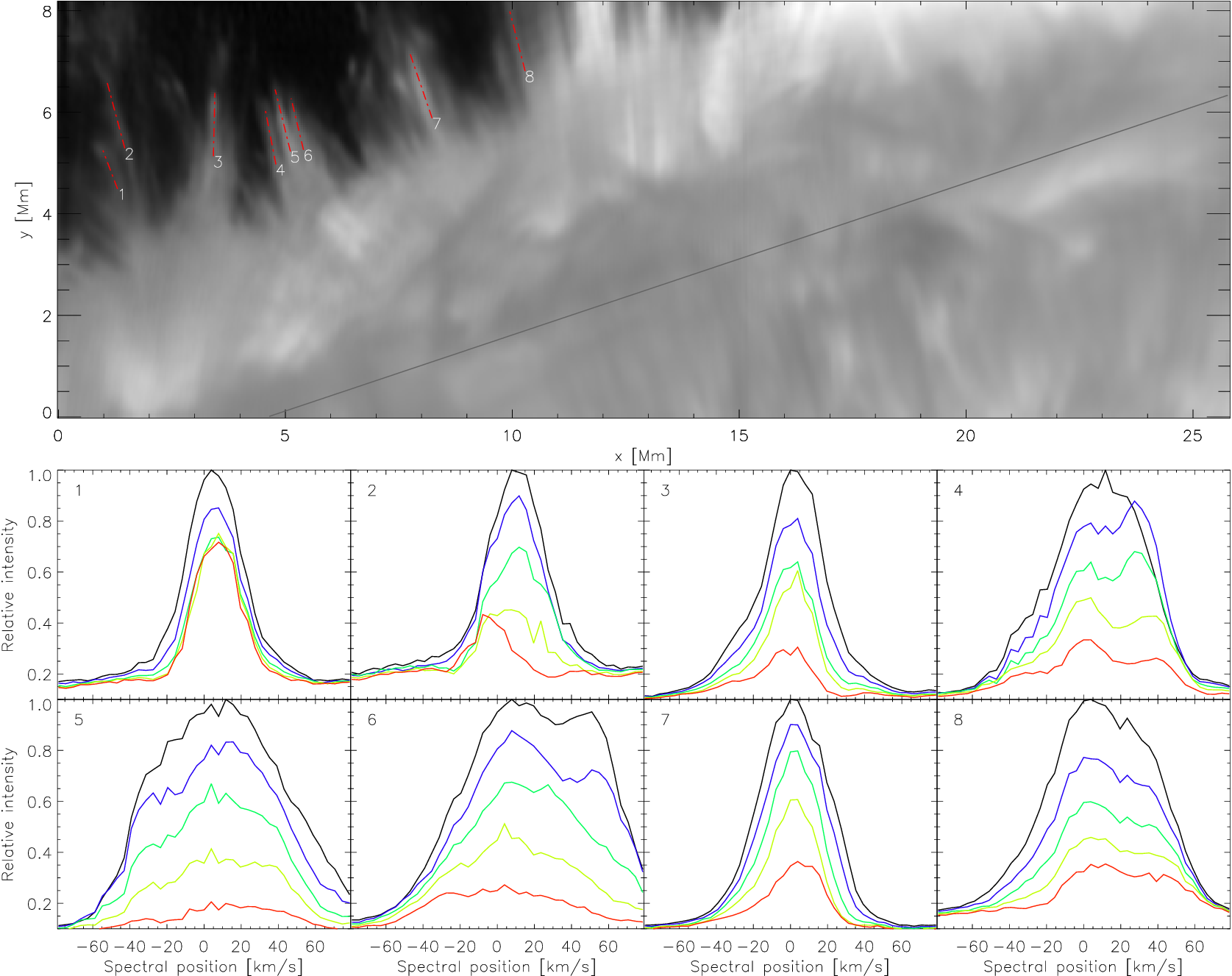}
  \caption{Spectral line profile as function of height for 8 spicules.
    The upper panel shows a cutout of the full FOV of line center \ha\ with the spicules studied marked by red dash-dot lines.
    The image is radial filtered to make the spicules stand out more clearly.
    The inclined gray line towards the lower right is the position of the photospheric limb.
    In the remaining panels the spectral line profiles are shown for 5 different heights with increments of 25\%.
    The black line is from the position closest to the limb while red is from the position furthest away from the limb.
    Each line profile is an average of three line profiles perpendicular to the spicule.
    The normalization factor is individual for each spicule.}
  \label{fig:asym}
\end{figure*}

The spatial scales across the field showing collective behavior is up to 2 Mm with a typical size of 500~km.
Examples of the size and number of spicules in a group can be seen in the top panel of Fig.~\ref{fig:asym}, where the spicules in immediate vicinity of spicule no.
3, 5 and 7 form three different groups of spicules displaying synchronous behavior.

If we study the spectral line profiles for individual spicules far away from the limb we observe that many line profiles are asymmetric and some even show multiple peaks indicating the presence of multiple components.
Fig.~\ref{fig:asym} show line profiles at various heights for eight different spicules.
The chosen spicules are located far away from the photospheric limb (gray line) and outside most of the (average) chromospheric limb.
This selection was an effort to reduce the line-of-sight superposition which is enormous close to the limb.
All spicules were selected by eye and the endpoints are arbitrary in the sense that no criterion was used to define the top and foot point.
Spicules no. 4, 5, 6 and 8 in particular show irregular line profiles which deviate significantly from a Gaussian profile, which is the expected line profile when thermal motion is dominating the line broadening.
Indications of multiple peaks are apparent in several examples and have been found in many more spicules not shown in this figure.
 
\ha\ is usually an optically thick line in the solar atmosphere, however that is not always the case off-limb.
We analyze spicules sufficiently far from the limb so the spectral lines generally have a Gaussian shape and we therefore treat spicules we study as being optically thin.

To analyze the observed asymmetrical line profiles we fit a double Gaussian line profile using asymmetry analysis following the work of \cite{de_pontieu_et_al_2009} and \cite{tian_et_al_2011}.
The following double Gaussian function is fitted:
\begin{equation}
  I(\lambda) = b + h_1 e^{-\frac{(\lambda - \mu_1)^2}{2\sigma_1^2}}+ h_2 e^{-\frac{(\lambda-\mu_2)^2}{2\sigma_2^2}},
\end{equation}
where $\lambda$ is the wavelength, $b$ is the background signal level, $h$ is the intensity, $\mu$ is the mean of the distribution and $\sigma$ is the standard deviation of the distribution.
The number in the subscript refers to the two Gaussians.
This function contains 7 free parameters: $b,h_1,\mu_1,\sigma_1, h_2, \mu_2$ and $\sigma_2$.

The asymmetry analysis involves computing the asymmetry of a line profile, which we define for a discrete intensity line profile $I(\lambda)$ to be:
\begin{equation}
  a(i) = \frac{\sum_{\lambda=\lambda_0' - i\Delta \lambda}^{\lambda_0'-i\Delta\lambda -\Delta\lambda'} I(\lambda) - \sum_{\lambda=\lambda_0' + i\Delta\lambda}^{{\lambda_0' + i\Delta\lambda+\Delta\lambda'}} I(\lambda)}{I_{max} -I_{min} },
\end{equation}
with $i \in [0,1,2,\ldots]$.
$\lambda_0'$ is the location of the peak of the line profile, $\Delta\lambda'$ and $\Delta\lambda$ are user defined intervals (both set to 10~km~s$^{-1}$) and $I_{max}$ and $I_{min}$ are the maximum and minimum value of the line profile respectively.
The location of maximum asymmetry is found from the peak of $|a(i)|$.

The procedure first counts the number of peaks in the line profile.
If there are two peaks in the line profile $\mu_1$ and $\mu_2$ are placed at one peak each.
If there only is one peak the asymmetry analysis is performed and the level of asymmetry is compared to a threshold.
The threshold was set to $0.2$ because at less asymmetry the double Gaussian fit would often fail.
If sufficient asymmetry is present we place one $\mu$ at the peak of the line profile and the other at the location of maximum asymmetry.
If the level of asymmetry is below the threshold then a single Gaussian function is fitted instead.
The $\mu$'s are allowed to have a small ``wiggle room'', $\pm 15$~km~s$^{-1}$, to achieve best fit.
This effectively reduces the number of free parameters down to 5.

For the asymmetric single-peaked line profiles we perform the fit 50 times while varying the starting estimate for the $\mu$ placed at the location of peak asymmetry.
We do this because the solution from the fitting procedure is at times dependent on the initial guesses for the parameters with that $\mu$ being a more sensitive parameter.
All unique solutions the procedure converges to are stored and near identical solutions are grouped as one and the number of groups are counted.
If there is more than one group, i.e. the solution to the fitting is ambiguous, the line profile in question is discarded from further analysis.

\begin{figure}[!t]
  \centering
  \includegraphics[width=\columnwidth]{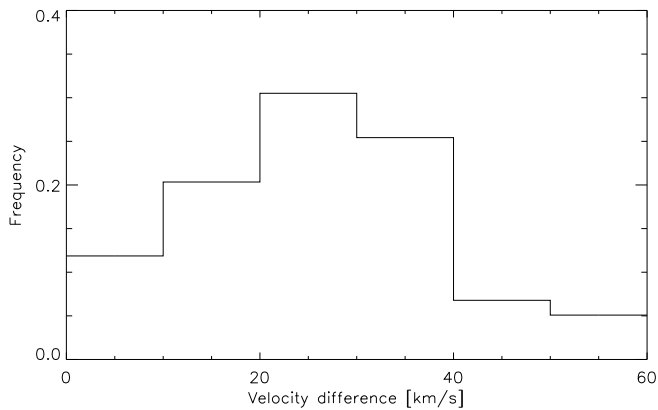}
  \caption{Histogram showing the average velocity difference between the two single Gaussian components for 57 spicules.}
  \label{fig:diffv}
\end{figure}

To quantify the asymmetric line profiles we selected 91 spicules by eye in moments with excellent seeing, in similar fashion as for Fig.~\ref{fig:asym}.
The line profiles constructed at each height were an average of the two closest pixels perpendicular to the spicule axis and the pixel on the spicule axis.
Double Gaussian fit was performed for each line profile and from this we computed the average velocity difference in each spicule, shown in Fig.~\ref{fig:diffv}.
At each height the level of asymmetry of the line profile was assessed.
Only heights where the level of asymmetry was above the threshold were included in the average.
In the histogram 57 spicules are included.
The remaining spicules either did not display sufficient asymmetry (19), the fitting result was ambiguous (10), or a combination of the two (3).

\begin{figure*}[!t]
  \includegraphics[width=\textwidth]{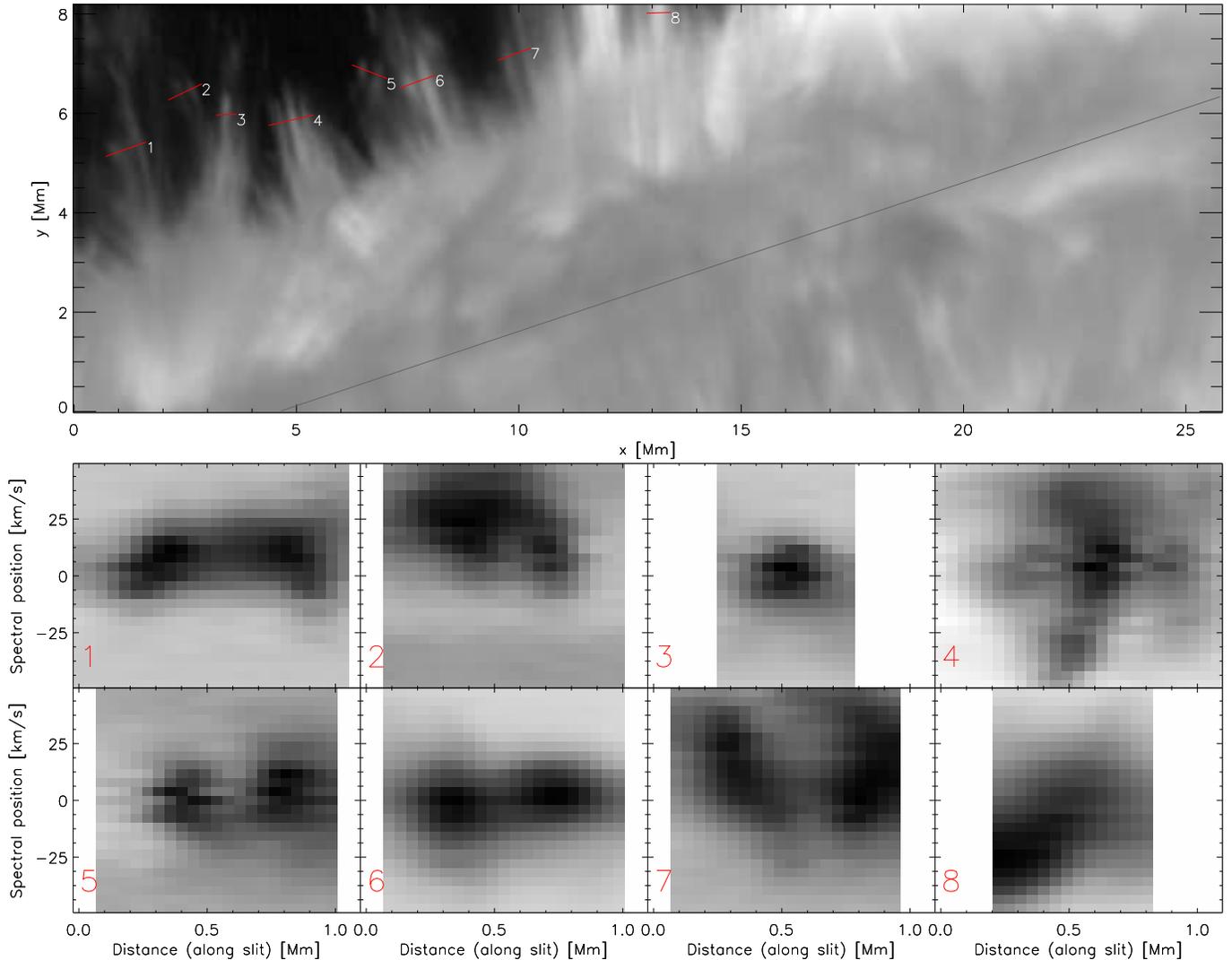}
  \caption{$\lambda-x$ slices across many spicules.
    The upper panel shows 8 artificial slits as red lines overplotted on a cutout from the full FOV in line center \ha.
    The inclined gray line towards the lower right is the position of the photospheric limb and the image is radial filtered.
    The remaining panels show the $\lambda-x$ slices in inverted color table with the distance measure proceeding from left to right along the slits in the upper panel.}
  \label{fig:lxslice}
\end{figure*}

Further complexity in spicules is apparent when we look at $\lambda-x$ slices across spicules, some examples in Fig.~\ref{fig:lxslice}.
The top panel shows the spatial position of the slits with corresponding $\lambda-x$ slices in the remaining panels.
For most examples a striking feature is the tilt of the spectral line as function of perpendicular distance across the spicules, most clearly seen in slit 8.
A tilt in a $\lambda-x$ slice indicates a velocity gradient across the feature and may be interpreted as a sign of rotational motion.
The central spicule in slit 4 is even more complex.
It shows two peaks in the red and blue wings indicating an overlap of two threads and both show different spectral tilts.
For slits no. 3 and 5 there are no clear velocity gradients across the spicules.

\section{Discussion}
We studied \ha\ spicules in an excellent SST off-limb time series and often observe group behavior: multiple spicules act as 'threads' in a larger structure, with each individual thread displaying a complex dynamic nature.
The overall apparent motion is synchronized with the other threads in the structure.
We find that spectral line profiles of isolated spicules far above the limb are often asymmetric, sometimes with clear double peaks, and are best fitted by a double Gaussian line profile.
The scenario that is emerging from these observations is that spicules are fundamentally multi-threaded with strong perpendicular flows inside 'one' spicule.
This multi-threaded nature is compatible with a whole volume undergoing spicular motions, with only a few threads visible at one time, and apparently harbouring turbulent perpendicular flows.
The perpendicular flows peak at around 20--30~km~s$^{-1}$ which is compatible with previous reports of Alfvenic waves.
All of this provide constraints on models for the formation and evolution of spicules.

\cite{suematsu_et_al_2008} suggest that more than 50\% of spicules appear as double-threaded structures in the Ca {\sc II} 3968 \AA\ passband of Hinode.
In our observations, we actually find that we often see more than two threads.
They report separation of the order of a few tens of an arc-second of the individual threads while we see synchronous behavior at larger spatial extents at times exceeding 2 arc-seconds.

We regard chance alignment of unrelated overlapping spicules as an unlikely cause of our findings: we select spicules far from the limb to avoid line-of-sight superposition and we often find near constant spectral asymmetry as function of height along the spicule.

Which type of spicules are we analyzing? The classification of spicules into type I and type II are traditionally made in the Ca {\sc II} 3968 \AA\ passband.
Here we analyze \ha\ data and it is unclear how these types of spicules manifest in this diagnostic - we are unaware of any detailed comparison of \ha\ and Ca~H spicules that addresses this question in the literature. \cite{mcintosh_et_al_2008} and \cite{pereira_2013_blur} show images of off-limb spicules in both Ca {\sc II} and \ha\ at the same time which show that the appearance is similar in the two spectral lines. 
We therefore assume the diagnostics are comparable for the task of classifying spicules.
\cite{de_pontieu_et_al_2007a} and \cite{pereira_2012} indicate that type I spicules are almost exclusively present in active regions.
The target of our data set is magnetically much less active compared to active regions and we observe very little downward motion of the spicules.
We therefore speculate that we observe type II spicules.

A possible explanation for the velocity gradient across spicules can be inferred from the work of \cite{ballegooijen_et_al_2011}.
These authors study the heating of the corona caused by Alfvenic turbulence.
They developed a numerical model where upward propagating waves are created from flux-tube foot-point motions.
Due to interaction between downward propagating reflected waves, turbulence is created in the flux-tube in the photosphere and chromosphere.
Our results show that the line-of-sight velocity across spicules can be complex, not necessarily a constant gradient across the spicule.
It is tempting to speculate whether we observe the Alfvenic turbulence proposed by \cite{ballegooijen_et_al_2011} but the speculation is limited by our spatial resolution ($\sim 130$~km) compared to the less-than 100~km spatial scales in the model.
The complexity we sometimes see in our $\lambda-x$ slices can be an indication of higher rotational modes than the fundamental mode which may provide support for this theory.

\cite{shelyag_et_al_2011} use numerical simulations to show that vorticity in the photosphere can be generated as a result of interaction between granular flows and the magnetic field in the intergranular lanes.
The simulation does not include the chromosphere and it is therefore difficult to say if the vorticity we observe is connected to the process they study.

What causes the multi-threaded nature of spicules?
A recent paper by \cite{antolin_et_al_2014} shows that the observed fine strand-like structure of coronal loops can be generated by a combination of the line-of-sight angle and vortices generated by the Kelvin-Helmholtz instability.
Transverse MHD waves generate the instability which creates the vortices.
The authors show intensity images in the hot Fe IX 171~\AA\ passband, and while not directly comparable to \ha\ the similarities are striking.
Further modeling is required to establish whether the mechanism can explain the multi-threaded nature of spicules we see in \ha.

\acknowledgements
The research leading to these results has received funding from both the Research Council of Norway and the European Research Council under the European Union's Seventh Framework Programme (FP7/2007-2013) / ERC grant agreement nr. 291058. The Swedish 1-m Solar Telescope is operated on the island of La Palma by the Institute for Solar Physics of Stockholm University in the Spanish Observatorio del Roque de los Muchachos of the Instituto de Astrof\' isica de Canarias.
B.D.P. is supported by NASA contract NNG09FA40C (IRIS), and NASA grants NNX11AN98G and NNM12AB40P.

%\bibliographystyle{apj}
%\bibliography{../ref}

\end{document}